%% file: main.tex
\documentclass{IEEEcsmag}

\usepackage[colorlinks,urlcolor=blue,linkcolor=blue,citecolor=blue]{hyperref}
\expandafter\def\expandafter\UrlBreaks\expandafter{\UrlBreaks\do\/\do\*\do\-\do\~\do\'\do\"\do\-}

\include{macro}

\jtitle{This manuscript is an extended version of the original paper accepted by IEEE Micro.}

\begin{document}

\title{From Block to Byte: Transforming PCIe SSDs with CXL Memory Protocol and Instruction Annotation}

\author{Miryeong Kwon$^{*}$, Donghyun Gouk$^{*}$, Junhyeok Jang$^{*}$, Jinwoo Baek$^{*}$, Hyunwoo You$^{*}$, Sangyoon Ji$^{*}$, Hongjoo Jung$^{*}$, Junseok Moon$^{*}$, Seungkwan Kang$^{\ddagger}$, Seungjun Lee$^{\ddagger}$, Myoungsoo Jung$^{*\dagger\ddagger}$
}
\affil
{
\\
\\$^{*}$Next-Generation Silicon and Research Division, \textbf{Panmnesia, Inc.}, Daejeon, South Korea
\\$^{\dagger}$Advanced Product Engineering Division, \textbf{Panmnesia, Inc.}, Seoul, South Korea
\\$^\ddagger$KAIST, Daejeon, South Korea}

\begin{abstract}\looseness-1 This paper explores how Compute Express Link (CXL) can transform PCIe-based block storage into a scalable, byte-addressable working memory. 
  We address the challenges of adapting block storage to CXL\textquotesingle{}s memory-centric model by emphasizing cacheability as a key enabler and advocating for Type 3 endpoint devices, referred to as CXL-SSDs. 
  To validate our approach, we prototype a CXL-SSD on a custom FPGA platform and propose annotation mechanisms, Determinism and Bufferability, to enhance performance while preserving data persistency. 
  Our simulation-based evaluation demonstrates that CXL-SSD achieves 10.9$\times$ better performance than PCIe-based memory expanders and further reduces latency by 5.4$\times$ with annotation enhancements. 
  In workloads with high locality, CXL-SSD approaches DRAM-like performance due to efficient on-chip caching. This work highlights the feasibility of integrating block storage into CXL\textquotesingle{}s ecosystem and provides a foundation for future memory-storage convergence.
  \vspace{-10pt}
\end{abstract}

\maketitle

\input{introduction.tex}

\section{CXL MEMORY PROTOCOL} \label{sec:background}
\input{background.tex}

\input{background1.tex}

\section{TRANSFORMATION OF PCIE SSD}
\input{background2.tex}

\input{design.tex}

\section{INSTRUCTION ANNOTATION} \label{sec:design_extension}
\input{design_extension.tex}

\section{EVALUATION}
\input{evaluation.tex}

\section{DISAGGREGATION DISCUSSION} \label{sec:design_disaggregation}

\input{design_disaggregation.tex}

\section{Conclusion} \label{sec:conclusion}
\input{conclusion.tex}

\section{ACKNOWLEDGEMENT}
\label{sec:acknowledgement}
\input{acknowledgements.tex}

\newpage

\bibliographystyle{ieeetr}
\bibliography{reference}

\end{document}

%% file: macro.tex
\usepackage{xcolor}
\usepackage{xspace}
\usepackage{subcaption}
\usepackage{pifont}
\usepackage{times}
\usepackage{graphbox}
\usepackage{graphicx}
\usepackage{epstopdf}
\usepackage{dblfloatfix} 
\usepackage[htt]{hyphenat}
\let\rfoot\relax
\usepackage{fancyhdr}
\usepackage{setspace}
\usepackage{relsize}
\usepackage{textcomp}
\usepackage{circledsteps}
\usepackage{url}
\usepackage[htt]{hyphenat}
\usepackage[super]{nth}
\usepackage{comment}
\usepackage{transparent}
\usepackage{nicematrix,booktabs}
\usepackage{textcomp}

\usepackage[T1]{fontenc}
\usepackage{lipsum}
\usepackage[normalem]{ulem}
\makeatletter
\def\SOUL@hlpreamble{%
\setul{\dimexpr\dp\strutbox-2pt}{\dimexpr\ht\strutbox+\dp\strutbox-2pt\relax}
\let\SOUL@stcolor\SOUL@hlcolor
\SOUL@stpreamble
}
\makeatother

\newcommand\khlc[1][yellow]{
  \bgroup
  \markoverwith{\textcolor{#1}{\rule[-.5ex]{1pt}{2.5ex}}}
  \ULon
}

\newcommand{\todo}[1]{\bgroup\color{white}\textbf{\khlc[black]{TODO: [#1]}}\egroup\xspace}
\newcommand{\fixme}[1]{\bgroup\color{red}\textbf{\khlc{FIXME: [#1]}}\egroup\xspace}
\newcommand{\pointer}[1]{\bgroup\color{white}\textbf{\khlc[red]{POINTER: [#1 is working here]}}\egroup\xspace}
\newcommand{\red}[1]{\bgroup\color{red}#1\egroup\xspace}
\newcommand{\blue}[1]{\bgroup\color{blue}#1\egroup\xspace}

\newcommand{\reviewer}[1]{\bgroup\color{blue}#1\egroup\xspace}
\newcommand{\ranswer}[1]{\bgroup\color{red}#1\egroup\xspace}

\let\labelindent\relax
\usepackage{enumitem}
\usepackage{amssymb}
\usepackage[framemethod=tikz]{mdframed}
\usepackage[numbers,sort&compress]{natbib}
\tikzset{
    vertical align/.style={
        baseline=-.5*(height("$+$")-depth("$+$"))
    }
}

\usepackage{calc}
\usepackage{amsmath,algorithm,algpseudocode}
\usepackage{cases}
\algrenewcommand\algorithmicindent{0.5em}
\makeatletter
\algrenewcommand\ALG@beginalgorithmic{\footnotesize}
\makeatother
\makeatletter
\renewcommand{\Function}[2]{%
  \csname ALG@cmd@\ALG@L @Function\endcsname{#1}{#2}%
  \def\jayden@currentfunction{#1}%
}
\newcommand{\funclabel}[1]{%
  \@bsphack
  \protected@write\@auxout{}{%
    \string\newlabel{#1}{{\jayden@currentfunction}{\thepage}}%
  }%
  \@esphack
}
\makeatother

\usepackage{pythonhighlight}
\definecolor{codegray}{rgb}{0.5,0.5,0.5}
\lstdefinestyle{pythonStyle}{
  basicstyle=\tiny\ttfamily\footnotesize\linespread{0.5},
  basewidth = {.54em},
  commentstyle=\color{codegray},
  frame=single,
  language=Python,
  stepnumber=1,
  numbers=left,
  numbersep=5pt,
  numberstyle=\tiny\color{codegray},
  tabsize=1,
  showspaces=false,
  showstringspaces=false,
  breaklines=false,
  mathescape,
  keywordstyle={\color{black}},
  emph={Load, GraphPre, BatchPre, Aggr, Trans},
  emphstyle={\bfseries\color{orange}},
  moredelim=**[is][\color{red}]{~}{~},
  moredelim=**[is][\color{blue}]{<}{>},
  moredelim=**[is][\color{orange}]{@}{@},
  literate={\\~}{{\textasciitilde}}1
  {\\<}{{\unichar{"003C}}}1
  {\\>}{{\unichar{"003E}}}1
  {\\@}{{\unichar{"0040}}}1
}

\usepackage{tabularx}
\usepackage{array}
\usepackage{multirow}
\usepackage{booktabs}
\usepackage[para]{threeparttable}
\usepackage{colortbl}
\usepackage{multirow}

\makeatletter
\def\hlinewd#1{%
\noalign{\ifnum0=`}\fi\hrule \@height #1 %
\futurelet\reserved@a\@xhline}
\makeatother
\usepackage{hhline}
\newcolumntype{C}{>{\centering\arraybackslash}X}

\usepackage[framemethod=tikz]{mdframed}
\usepackage{marginnote}
\usepackage{pdfcomment}
\mdfsetup{
  hidealllines=true,
  innerleftmargin=2pt,
  innerrightmargin=2pt,
  innertopmargin=2pt,
  innerbottommargin=2pt,
  leftmargin=-2pt,
  rightmargin=-2pt,
  skipabove=-2pt,
  skipbelow=-2pt,
  backgroundcolor=blue!20}

\newlength{\markerHeight}
\newlength{\markerMargin}
\newlength{\linespace}
\newlength{\linedepth}

\sbox0{yf}
\definecolor{mylime}{RGB}{205, 220, 57}
\definecolor{mygreen}{RGB}{60, 200, 0}
\definecolor{myblue}{RGB}{0, 51, 204}

\colorlet{soulred}{red!20}
\colorlet{soulgreen}{green!20}
\colorlet{soulblue}{blue!20}

\usepackage{enumitem}
\setlistdepth{5}
\renewlist{itemize}{itemize}{5}
\setlist[itemize,1]{label=$\bullet$}
\setlist[itemize,2]{label=$\circ$}
\setlist[itemize,3]{label=$\ast$}
\setlist[itemize,4]{label=-}
\setlist[itemize,5]{label=$\cdot$}

\makeatletter
\def\SOUL@hlpreamble{%
\setul{\dimexpr\dp\strutbox-2pt}{\dimexpr\ht\strutbox+\dp\strutbox-2pt\relax}
\let\SOUL@stcolor\SOUL@hlcolor
\SOUL@stpreamble
}
\makeatother

\definecolor{reviewerA_bg}{RGB}{253, 255, 117}
\definecolor{reviewerB_bg}{RGB}{199, 253, 163}
\definecolor{reviewerC_bg}{RGB}{168, 254, 237}
\definecolor{reviewerD_bg}{RGB}{170, 222, 255}
\definecolor{reviewerE_bg}{RGB}{221, 170, 255}
\definecolor{reviewerF_bg}{RGB}{255, 185, 185}
\definecolor{reviewerA_bg_l}{RGB}{254, 255, 198}
\definecolor{reviewerB_bg_l}{RGB}{228, 255, 210}
\definecolor{reviewerC_bg_l}{RGB}{203, 255, 245}
\definecolor{reviewerD_bg_l}{RGB}{205, 236, 255}
\definecolor{reviewerE_bg_l}{RGB}{235, 204, 255}
\definecolor{reviewerF_bg_l}{RGB}{255, 219, 225}
\definecolor{reviewerDel_bg}{RGB}{81, 81, 81}

%% file: introduction.tex
\chapteri{C}ache coherence interconnects have recently emerged to integrate CPUs, accelerators, and memory components into a unified, heterogeneous computing domain. 
These interconnect technologies ensure data coherency between CPU memory and device-attached private memory, creating a new paradigm of globally shared memory and network space. 
Among several efforts to establish such connectivity, including Gen-Z \cite{genzconsortium} and Cache coherent interconnect for accelerators (CCIX) \cite{ccixconsortium}, \emph{Compute Express Link} (CXL) has become the first open interconnect protocol capable of supporting diverse processors and device endpoints. 
With the absorption of Gen-Z, CXL stands out as a promising interconnect interface due to its high-speed coherence control and seamless compatibility with the widely adopted PCIe standard. 
This makes it particularly advantageous for a wide range of datacenter-scale hardware, including CPUs, GPUs, FPGAs, and domain-specific ASICs. 
Furthermore, the CXL consortium has highlighted its potential for memory disaggregation, enabling pooling of DRAM and byte-addressable persistent memory.

Despite its versatility in handling computing resources and memory components, CXL currently excludes block storage, leaving open the question of whether storage devices can leverage the benefits of CXL. 
Storage designers and system architects may ask: what specific advantages can CXL offer to block storage, and why is this integration meaningful? 
Although a few industrial conceptual prototypes exist \cite{zhong2024managing}, the practical benefits of CXL over PCIe in this context remain uncertain. 
Notably, all current versions of CXL utilize PCIe\textquotesingle{}s physical media and coding sublayers \cite{cxlspec}, resulting in identical analog characteristics and low-level performance between the two interfaces. 
In addition, as CXL is designed to unify various hardware accelerators and computing devices into a coherent memory pool, careful consideration is required to determine the appropriate role of block storage within this ecosystem.

In this paper, we argue that CXL can effectively transform PCIe-based block storage into a large, scalable working memory by addressing the key questions outlined earlier. We propose that CXL is a cost-effective and practical interconnect technology capable of bridging the block-based semantics of PCIe storage to the memory-compatible, byte-addressable semantics required for efficient operation as working memory. Achieving this requires careful integration of block storage into the CXL interconnect network, taking into account the diversity of device types and protocols supported by CXL.

This paper begins by examining the limitations that make PCIe storage impractical for use as a memory expander (cf. the Preliminary Performance Model of CXL-SSD section). 
While both PCIe and CXL share the same physical layer and allow access through memory instructions (e.g., loads and stores), we emphasize that CXL introduces a critical characteristic for inclusion in the memory hierarchy: cacheability. 
This feature enables on-chip cache hits to bypass access to the underlying memory, offering a significant advantage over PCIe. 
Such cacheability is essential for improving memory access latency and efficiency when integrating block storage into a coherent memory system. 
We advocate for the adoption of Type 3 endpoint devices as the optimal choice for implementing flash-based CXL memory devices, referred to as CXL-SSDs, instead of Type 2 devices. 
To substantiate this argument, we prototype a CXL-SSD by incorporating essential CXL intellectual properties (IPs) into a custom FPGA platform configured as 16nm. 
This prototype serves as a foundational step in demonstrating the potential of CXL-enabled block storage as a scalable and efficient memory expander.

Although CXL represents the most promising interface for bringing block storage closer to the CPU, the projected performance of CXL-SSD remains significantly behind that of DRAM-based media. To address this gap, we propose instruction annotation mechanisms: (i) Determinism and (ii) Bufferability. These mechanisms are designed to enhance the performance of CXL-SSD while preserving data persistency as block storage. Since the effectiveness of these annotation instructions depends on their application context, this paper also explores several use cases where CXL-SSD benefits from their adoption.

To investigate the design space, we develop a full-system simulation model based on the CXL-SSD prototype. 
Our simulation-based evaluation demonstrates that CXL-SSD achieves 10.9$\times$ better performance compared to a PCIe-based memory expander utilizing the same underlying flash media. 
Furthermore, the annotation-augmented CXL-SSD reduces latency by an additional 5.4$\times$, on average, relative to the original CXL-SSD. 
In scenarios with high locality during application execution, CXL-SSD achieves performance levels comparable to DRAM-based local memory, primarily due to on-chip cache hits.

%% file: background.tex
\begin{figure}
    \begin{subfigure}{\linewidth}
        \vspace{0pt}
        \centering
        \includegraphics[width=1\linewidth]{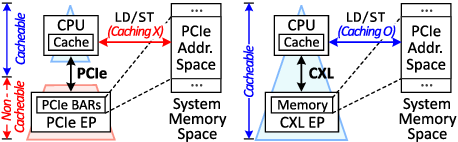}
        \begin{tabularx}{\textwidth}{
            p{\dimexpr.49\linewidth-2\tabcolsep-1.3333\arrayrulewidth}
            p{\dimexpr.49\linewidth-2\tabcolsep-1.3333\arrayrulewidth}
            }
              \vspace{-5pt} \caption{PCIe-based expansion.} \label{fig:pcie-cxl-non-cacheable}
            & \vspace{-5pt} \caption{CXL-based expansion.} \label{fig:pcie-cxl-cacheable}         
        \end{tabularx}
    \end{subfigure}
    \vspace{-3pt}
    \caption{Memory expansion based on PCIe and CXL.} \label{fig:pcie-cxl}
    \vspace{3pt}
  \end{figure}

\subsection{Why CXL Memory for PCIe Storage?}
\noindent \textbf{Byte-addressability.} Achieving byte-addressability for PCIe storage and integrating it into working memory devices has long been a key goal. For instance, industry prototypes and the NVMe standard provide byte-addressability by exposing an SSD\textquotesingle{}s internal memory or buffer to PCIe base address registers (BARs). Since BARs can be directly mapped to the system memory space, host-side kernels and applications can access these exposed memory or buffer resources like local memory, using load/store instructions instead of block-level operations. To mitigate the long latencies associated with the backend block media of SSDs (e.g., Z-NAND, Flash, Optane SSD), the internal memory or buffer can serve as a write-back inclusive cache for the backend storage, effectively enhancing access performance.

\noindent \textbf{Limits with non-cacheable accesses.} While PCIe bandwidth is sufficient to support far memory (e.g., 63 GB/s to 121 GB/s for Gen5/6 with 16 lanes), PCIe treats block storage devices as peripheral components managed and accessed by the host-side CPU. As shown in Figure \ref{fig:pcie-cxl-non-cacheable}, although storage devices can therefore process load/store requests through PCIe\textquotesingle{}s BARs, they are inherently limited in their ability to function as working memory in real-world systems. Specifically, memory-mapped BARs serve only as an interface for the host to communicate requests or control signals to the storage, requiring all load/store requests from the CPU to be non-cacheable and directly accessible.

This non-cacheable behavior significantly degrades the performance of memory accesses targeting BARs. If the CPU were to cache or buffer memory requests targeting the PCIe address space, the PCIe storage device would have no mechanism to detect or respond to these requests, potentially leading to critical issues such as system failure or storage disconnection. To avoid such scenarios, x86 instruction set architectures from both Intel and AMD enforce restrictions that prevent caching of PCIe-related memory requests at the CPU level. Unfortunately, this restriction excludes storage-integrated memory expanders from the conventional memory hierarchy, denying them the performance benefits of CPU caching.

\noindent \textbf{Compute express link.} CXL is a cache-coherent interconnect technology originally designed to support a wide range of accelerators and memory devices. It enables one or more memory address spaces within the PCIe network domain to be accessed coherently by multiple processors and hardware accelerators, leveraging a multi-protocol approach. This multi-protocol capability extends beyond the I/O semantics of the traditional PCIe interface, ensuring that all CXL device types remain compatible with most existing PCIe devices, including SSDs. Details of the CXL device types will be discussed in the Multi-Protocol and Device Type Classification section.

Although CXL is built on PCIe, it fundamentally ensures cache coherence across all computing complexes within the same CXL hierarchy. As shown in Figure \ref{fig:pcie-cxl-cacheable}, this allows load/store requests targeting the PCIe address space to be cacheable, addressing one of the critical limitations of PCIe. While current CXL implementations primarily focus on DRAM and PMEM for memory pooling, we argue that CXL can transform PCIe storage by enabling a memory-like, byte-addressable interface. By integrating PCIe storage into its cache-coherent memory space through its multi-protocol capability, CXL has the potential to create a significantly larger memory pool compared to traditional DRAM- or PMEM-based memory expansion technologies.

%% file: background1.tex
\begin{figure}
  \begin{subfigure}{\linewidth}
      \vspace{0pt}
      \centering
      \includegraphics[width=1\linewidth]{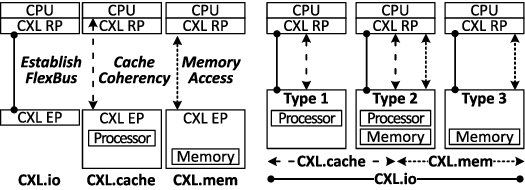}
      \begin{tabularx}{\textwidth}{
          p{\dimexpr.49\linewidth-2\tabcolsep-1.3333\arrayrulewidth}
          p{\dimexpr.49\linewidth-2\tabcolsep-1.3333\arrayrulewidth}
          }
            \vspace{-10pt} \caption{Subprotocols.} \label{fig:cxl-subprotocol}
          & \vspace{-10pt} \caption{Types of EPs.} \label{fig:cxl-ep-types}         
      \end{tabularx}
  \end{subfigure}
  \vspace{-5pt}
  \caption{CXL subprotocols and EP types.} \label{fig:cxl}
  \vspace{5pt}
\end{figure}

\subsection{Multi-Protocol and Device Type Classification}
As shown in Figure \ref{fig:cxl-subprotocol}, CXL defines three sub-protocols, (i) CXL.io, (ii) CXL.mem, and (iii) CXL.cache, which correspond to three types of CXL devices, categorized as Type 1, Type 2, and Type 3 endpoints.

\noindent \textbf{Multi-protocol.} CXL.io serves as the foundational protocol required for communication between all CXL-attached devices and host CPUs. It extends the full functionality of PCIe, acting as a non-coherent load/store interface for I/O operations, such as device discovery, enumeration, and host address configuration. To achieve this, CXL.io modifies PCIe\textquotesingle{}s hierarchical communication layers and establishes a high-speed I/O channel called FlexBus. FlexBus translates received CXL data into a format compatible with PCIe\textquotesingle{}s physical layers, including transaction, data, and link layers.

In contrast, CXL.cache and CXL.mem enhance FlexBus by introducing coherent cache and memory access capabilities. These protocols enable support for multiple device domains and facilitate remote memory management. \emph{CXL\textquotesingle{}s root port} (CXL RP) allows memory addresses exposed by underlying CXL devices to be mapped into a host\textquotesingle{}s cacheable system memory space. While this capability is primarily designed to unify diverse memory devices into a single, coherent memory pool, it can also be adapted to enable memory expanders leveraging various storage technologies.

\noindent\textbf{CXL device types.} CXL defines three device types -- Type 1, Type 2, and Type 3 -- based on how the multi-protocol features of CXL are utilized. Figure \ref{fig:cxl-ep-types} shows these device types along with the protocols each type employs.

Type 1 devices are equipped with a local cache but do not include internal DRAM components. These devices are particularly valuable for endpoint applications that require fully coherent cache access. Type 1 devices actively access the host\textquotesingle{}s memory data through their own cache, ensuring coherence with the host memory. This makes them suitable for domain-specific accelerators used in computationally intensive tasks, such as tensor processing units. Type 1 devices rely on CXL.cache and CXL.io protocols to provide full cache coherence capabilities.

Type 2 devices are designed for discrete accelerators equipped with high-performance memory modules, referred to in CXL as \emph{host-managed device memory} (HDM). By default, communication between the host and Type 2 devices is facilitated through CXL.io over PCIe. In addition, CXL.cache enables the device to access host-side memory coherently, while CXL.mem allows the host to access HDM directly. HDM differs fundamentally from the private memory modules used in conventional accelerators, such as GPUs. For example, while a host can access GPU memory (e.g., GDDR), this typically requires legacy memory copy operations. In contrast, a CXL-enabled host can manage HDM using cache-coherent load/store instructions, eliminating the need for explicit memory copies. Moreover, Type 2 devices can actively access the host\textquotesingle{}s CPU memory, leveraging the full range of features provided by CXL\textquotesingle{}s multi-protocol architecture.

Type 3 devices are designed for non-acceleration purposes, consisting solely of HDM without any processing components. These devices primarily operate using CXL.mem to handle load/store requests issued by the host, as CXL does not permit Type 3 devices to initiate requests to the host via CXL.cache. While Type 3 devices do not utilize CXL.cache, they are well-suited for host memory expansion, as CXL.mem provides fundamental read and write interfaces for HDM. Details on how HDM is used will be discussed in the
Integrating CXL Protocol into Block Storage section. In addition, CXL allows Type 3 devices to manage CXL.io on the device side, offering flexibility to support various I/O-specific demands.

%% file: background2.tex
\subsection{Integrating CXL Protocol into Block Storage}

\noindent \textbf{Device type consideration.} PCIe storage is inherently more than a simple, passive device. Beyond its backend block media, it incorporates substantial internal DRAM for buffering and caching incoming requests and associated data. In addition, it includes computational capabilities to perform tasks such as address translation and reliability management. 
Type 2 could be a viable option for PCIe storage, as they allow the utilization of HDM and the integration of data processing functionalities while maintaining awareness of host CPU semantics. 
Several industrial proposals have explored the use of CXL memory expanders as data processing units classified as Type 2 \cite{hermes2024udon}.

However, in this paper, we advocate for Type 3 devices as the optimal choice for implementing storage-integrated memory expanders in CXL. This approach simplifies the design by focusing on memory expansion capabilities while leveraging CXL.mem for efficient host-side access.

There are three reasons why we believe that Type 3 devices are better than Type 2 devices for the storage-integrated memory expander (CXL-SSD). First, even though Type 2 allows the host to handle the storage-side HDM directly, Type 2 is designed for computationally intensive applications. Because of this, only one device (per CXL RP) can be connected to a host system in CXL 2.0. While this has been extended in CXL 3.0, it is still limited to 16 per RP in cases where cache coherent is fully managed. This makes Type 2 devices not scalable as Type 3 devices can do (4,095 devices per RP). Second, incorporating full CXL.cache and CXL.mem functionality into CXL-SSDs can impose additional communication overhead, ultimately degrading overall performance. Specifically, each load/store request must verify the cache states within the PCIe storage\textquotesingle{}s computing complex, resulting in multiple CXL transactions for every I/O operation. While efficient management of the internal DRAM in PCIe storage is important, there is no need to coherently manage the host CPU\textquotesingle{}s caches, which adds unnecessary complexity.
Starting from CXL 3.0, the standard supports CXL switches, enabling a host system to connect multiple CXL memory expansion devices through multi-level switches. Such a switch hierarchy allows expansion beyond the maximum number of ports (approximately 4$\sim$8) provided by a single switch, enabling scalability up to the maximum memory capacity supported by CXL, which is 4PB.

Third, integrating PCIe storage as a Type 2 device requires the storage-side computing resources to request permission from the host whenever accessing its memory. This is because Type 2 devices must maintain full cache coherence between the host\textquotesingle{}s local memory and HDM, further reducing device-level I/O performance. Although a bias mode exists for accessing specific memory regions without explicit permission, when the goal is to focus entirely on memory operations, Type 2 becomes an expensive endpoint with limited scalability.

\noindent \textbf{Storage-side modification.} PCIe storage devices typically utilize a PCIe endpoint and NVMe controllers to parse incoming requests and facilitate data transfers between the host and the SSD\textquotesingle{}s internal DRAM. Consequently, the hardware changes required on the storage side to support CXL Type 3 are relatively minor, enabling most storage devices to adopt Type 3 with minimal modifications. For instance, a CXL storage controller can be implemented to manage CXL transaction packet formatting and CXL.io control by extending the existing PCIe endpoint logic. Similarly, the capabilities of the current NVMe controller, such as command parsing and page memory copying, can be simplified to handle the read and write interfaces of CXL.mem. The NVMe specification allows controllers to be implemented in either firmware or hardware. However, we recommend automating CXL.mem\textquotesingle{}s read and write operations in hardware while reserving firmware for managing internal DRAM and backend block media for traditional block accesses.

This approach is essential because CXL-SSDs operate within the memory hierarchy, unlike conventional storage that interacts through file systems. The slower performance imposed by firmware would directly degrade the host\textquotesingle{}s ability to handle memory instructions (e.g., load/store), which is far more critical in this context than traditional block-level read or write operations.

\begin{figure}
    \vspace{0pt}
    \centering
    \includegraphics[width=0.9\linewidth]{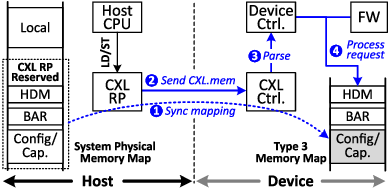}
    \vspace{10pt}
    \caption{System memory mapping.}
    \label{fig:background2-storage}
    \vspace{0pt}
\end{figure}

\noindent\textbf{System integration.} Figure \ref{fig:background2-storage} shows how CXL connects a PCIe storage device to a host and demonstrates how host-side users can directly access the storage device using load/store instructions. In this setup, the host\textquotesingle{}s system bus employs a CXL RP to connect a PCIe storage device configured as Type 3. A system option for disaggregating multiple storage devices from host resources will be discussed in the DISAGGREGATION DISCUSSION section.

During the host boot process, it enumerates the CXL devices connected to its RP and initializes them by mapping their internal memory spaces into the system memory. Specifically, the host retrieves the size of the CXL BAR and HDM from the PCIe storage devices and maps them to reserved system memory regions in the CXL RP. Notably, HDM is mapped to a cacheable memory address space, enabling users to access it through load/store instructions. Since the addresses mapped to CXL BAR and HDM differ from those initially managed by the Type 3 device, the CXL RP must notify the underlying CXL controller of their new mappings [\ding{182}]. This address space synchronization is achieved by writing the updated address information (e.g., remapped address offsets) to the CXL capability/configuration areas of the target storage device. When an application performs a load or store operation on the system memory mapped to HDM, the CXL RP generates a message, referred to as a CXL flit, and sends it to the target\textquotesingle{}s CXL storage controller via CXL.mem [\ding{183}]. The storage controller, in conjunction with the endpoint and CXL controllers, parses the flit to extract request details such as the command type and target address [\ding{184}]. Using this information, the controllers process the request by coordinating with the underlying storage firmware [\ding{185}].

%% file: design.tex
\begin{figure}
	\centering
    \begin{minipage}{.39\linewidth}
        \includegraphics[width=\linewidth]{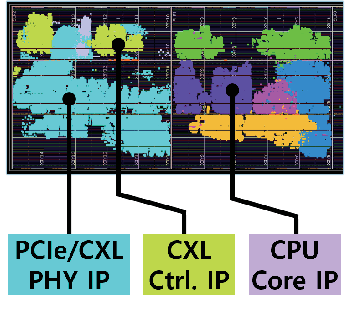}
    \end{minipage}
	\begin{minipage}{.59\linewidth}
		\includegraphics[width=\linewidth]{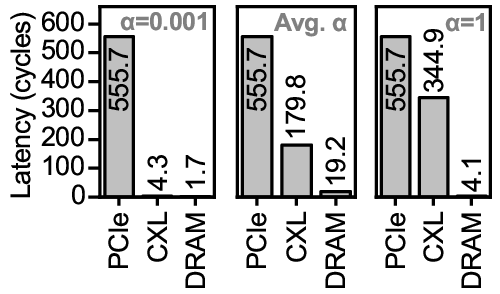}
	\end{minipage}
	\begin{subfigure}{1\linewidth}
		\centering
		\renewcommand*{\arraystretch}{1.2}
		\begin{tabularx}{\textwidth}{
            p{\dimexpr.39\linewidth-2\tabcolsep-1.3333\arrayrulewidth}
            p{\dimexpr.59\linewidth-2\tabcolsep-1.3333\arrayrulewidth}
            }
              \vspace{-5pt} \caption{Prototye.} \label{fig:floorplan}
            & \vspace{-5pt} \caption{Result.} \label{fig:motiv}
		\end{tabularx}
	\end{subfigure}
	\vspace{0pt}
    \caption{Performance of different memory systems.} \label{fig:perf_idff_memsys}
	\vspace{5pt}
  \end{figure}

\subsection{Preliminary Performance Model of CXL-SSD}
\noindent\textbf{Prototypes.}
As there are currently no commercial products supporting CXL for both processing complexes and endpoints in a modifiable setup, we prototype a CXL-enabled CPU and CXL storage to represent a host and a storage-integrated memory expander, respectively. The prototypes are implemented on two separate custom FPGA boards connected via a tailored PCIe backplane.

Figure \ref{fig:floorplan} shows the floorplan developed at the register-transfer level. For the host node, we integrate CXL.mem and CXL.io agents into an in-house RISC-V CPU featuring a 64-bit O3 dual-core architecture with 128KB L1 and 4MB L2 caches. The storage node utilizes a 32GB OpenExpress-based NVMe storage design implemented on a 16nm FPGA. The OpenExpress backend media emulates Z-NAND while buffering incoming CXL requests in internal DRAM. In addition to the CXL prototype (\texttt{CXL}), we evaluate a local DRAM-only system (\texttt{DRAM}) and a PCIe-based memory expander (\texttt{PCIe}). Both \texttt{PCIe} and \texttt{CXL} setups share the same backend storage, but their RP addresses are mapped to different regions of the host\textquotesingle{}s system memory.

\noindent \textbf{Latency impact of PCIe cacheable region.} We evaluate the latency impact using \emph{Apex-Map} \cite{strohmaier2005apex}, a global memory access benchmark designed for large-scale computing. 
This benchmark enables testing of the underlying memory system with varying levels of access locality and request sizes, controlled by the parameter $\alpha$. 
For our experiments, we set the request size to 64B, matching the last-level cache line size of our CPU. 
To focus on performance projection, we exclude time-consuming internal tasks such as garbage collection; 
these aspects and how \texttt{CXL} mitigates their latency will be discussed in the INSTRUCTION ANNOTATION section. 
Apex-Map generates 512 million synthetic memory instructions, varying $\alpha$ from 0.001 (highest locality) to 1 (lowest locality).

Figure \ref{fig:motiv} presents the latency (measured in CPU cycles) for the best case ($\alpha = 0.001$), average case ($0.001 \leq \alpha \leq 1$), and worst case ($\alpha = 1$). The best-case scenario highlights the significant advantage of \texttt{CXL} over PCIe-based memory expanders. While most memory requests in this scenario benefit from CPU cache hits, \texttt{PCIe} cannot leverage the host CPU caches, resulting in 129.5$\times$ longer latency compared to \texttt{CXL}. In contrast, \texttt{CXL} takes full advantage of CPU caches, achieving latency levels comparable to DRAM, demonstrating its superior efficiency.

\texttt{CXL} demonstrates a 3.1$\times$ improvement in performance over \texttt{PCIe} in the average case. 
Although the average-case performance of \texttt{CXL} is 9.3$\times$ slower than DRAM, this is a reasonable trade-off given that \texttt{CXL} leverages block storage as its underlying media. 
In the worst-case scenario, where no locality exists, \texttt{CXL} is unable to mitigate the inherent Z-NAND latency due to the fully random access patterns imposed by the benchmark, resulting in performance that is 84.1$\times$ slower than DRAM. Nevertheless, \texttt{CXL} still outperforms \texttt{PCIe} by 1.6$\times$ in this scenario, as it avoids the fully synchronized handling of memory requests required by PCIe BAR.

While the worst-case latency characteristics of \texttt{CXL} fall significantly short of DRAM-like behavior, it is important to note that most workloads exhibit high locality, with exceptions such as graph processing. Considering the substantial capacity provided by storage-integrated memory expanders, we posit that \texttt{CXL} can deliver significant benefits for a wide range of applications. Moreover, there is potential to optimize the observed long latencies by effectively leveraging the PCIe storage\textquotesingle{}s internal DRAM and backend block media, as discussed in the Why CXL Memory for PCIe Storage? section.

%% file: design_extension.tex
\begin{figure*}[]
    \hspace{-6pt}
    \begin{minipage}{0.5\linewidth}
        \centering
        \includegraphics[width=0.95\linewidth]{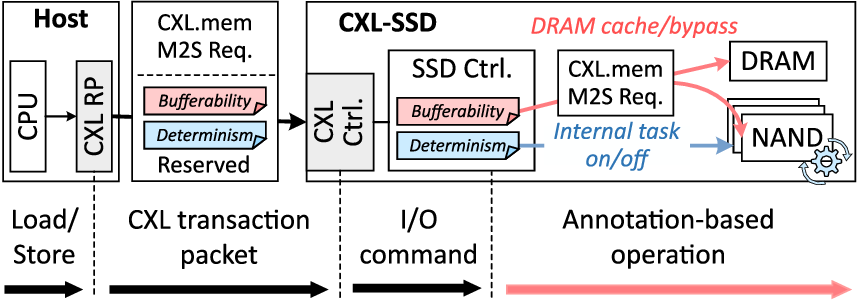}
        \vspace{4pt}
        \subcaption{Annotation workflow.}
        \label{fig:design_extention-annotation}
    \end{minipage}
    \hspace{6pt}
    \begin{minipage}{0.5\textwidth}
        \centering
        \includegraphics[width=0.95\linewidth]{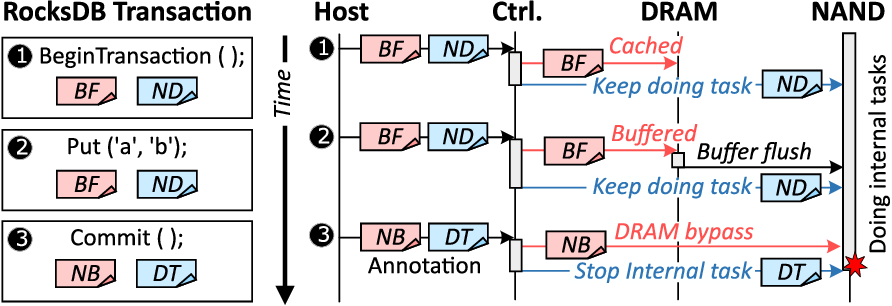}
        \vspace{4pt}
        \subcaption{RocksDB annotation example.}
        \label{fig:design_extention-example}
    \end{minipage}
    \vspace{20pt}
    \caption{Instruction annotation.}
    \vspace{5pt}
\end{figure*}

As CXL Type 3 is primarily designed for memory pooling rather than block storage, two critical challenges arise that warrant further discussion: (i) latency fluctuation and (ii) data persistence.

CXL.mem and CXL.io do not enforce strict management of load/store turnaround times, as CXL memory requests can be handled asynchronously. However, prolonged latency remains undesirable, as it can degrade overall host performance. For instance, in our performance projections, the PCIe storage device assumes the absence of internal tasks (cf. the Preliminary Performance Model of CXL-SSD section). In practice, the latency associated with these internal tasks is variable, influenced by firmware operations, and can significantly worsen responsiveness compared to baseline expectations. Moreover, when host-side libraries such as PMDK require strict data persistence, the current flushing mechanisms provided by CXL may be insufficient for underlying PCIe storage. CXL implements a \emph{global persistent flush} (GPF) register to ensure all data in the CXL network and SSD internal DRAMs is immediately written back to the backend media. However, this process can introduce additional latency, further impacting the performance consistency of storage-integrated memory expanders.

To address these issues, we propose two annotation features: (i) determinism and (ii) bufferability. These annotations can be embedded in CXL messages to convey host-level semantics to the underlying CXL controllers. CXL Type 3 already accommodates diverse I/O-specific demands via CXL.io (cf. the Multi-Protocol and Device Type Classification section), and CXL.mem includes a reserved field that can support such annotations. Specifically, CXL.mem defines the host and Type 3 as Master and Subordinate, respectively. Based on the direction of the request, CXL.mem messages are categorized as Master-to-Subordinate (M2S) or Subordinate-to-Master (S2M). These categories are further divided into subtypes: Request-with-Data (M2S Req), Request-without-Data (M2S RwD), No-Data-Response (S2M NDR), and Data-Response (S2M DRS).
The proposed annotations are designed to minimize overhead and do not modify the data payload itself. Instead, they are incorporated into M2S Req and S2M NDR messages, utilizing the 10-bit reserved fields available at the end of each message \cite{cxlspec}. 
This approach allows the annotations to be seamlessly included alongside the actual messages without incurring additional transmission costs.

\begin{table*}[]
  \begin{minipage}[t]{0.39\linewidth}
    \vspace{0pt}
    \centering
    \setlength\tabcolsep{3pt}
    \setlength\extrarowheight{1pt}
    \resizebox{1.0\linewidth}{!}{%
      \begin{tabular}{|c|l|}
        \hline
        \rowcolor{lightgray!20}
        \textbf{Param.} & \multicolumn{1}{c|}{\textbf{Value}}                                                \\ \hline
        CPU             & 4 GHz, 4 cores, 64-entry LSQ                                                       \\ \hline
        L1 cache        & \begin{tabular}[c]{@{}l@{}} I-cache: 64 KB, 8-way, 8-entry MSHR, 4-cycle \\ [-2pt]
                                D-cache: 64 KB, 12-way, 16-entry MSHR, 5-cycle\end{tabular} \\ \hline
        L2 cache        & 2 MB, 8-way, 32-entry, MSHR, 20-cycle                                              \\ \hline
        DRAM            & \begin{tabular}[c]{@{}l@{}}tRP=tRCD=tCAS=12.5ns \\[-2pt]
                                2 channels, 8 ranks, 8 banks\end{tabular}                           \\ \hline
        CXL             & \begin{tabular}[c]{@{}l@{}}CXL 3.1, PCIe 6.0, 64 GT/s, x4/x8 bifurcation \\[-2pt]
                                Endpoint device type: Type 3\end{tabular}   \\ \hline
        SSD             & \begin{tabular}[c]{@{}l@{}}DRAM cache: tRP=tRCD=9.1ns, tRAS=19ns \\[-2pt]
                                NAND flash: Z-NAND, tR=3us, tPROG=100us, tBERS=1ms\end{tabular}        \\ \hline
      \end{tabular}
    }
    \subcaption{Simulation setup.} \label{tab:setup}
  \end{minipage}
  \hfill
  \begin{minipage}[t]{0.6\linewidth}
    \vspace{0pt}
    \centering
    \setlength\tabcolsep{3pt}
    \resizebox{1.0\linewidth}{!}{
      \begin{tabular}{|c|l|rrrrr|c|l|rrrrr|}
        \hline
        \rowcolor{lightgray!20}
        \multicolumn{1}{|c|}{\textbf{cache}} &                                    &                                    & \multicolumn{1}{c}{\textbf{load}}  & \multicolumn{1}{c}{\textbf{store}} & \multicolumn{1}{c}{\textbf{Foot.}} & \multicolumn{1}{c|}{\textbf{LLC}} & \multicolumn{1}{|c|}{\textbf{cache}} &                                    &                                    & \multicolumn{1}{c}{\textbf{load}}  & \multicolumn{1}{c}{\textbf{store}} & \multicolumn{1}{c}{\textbf{Foot.}} & \multicolumn{1}{c|}{\textbf{LLC}} \\[-3pt]
        \rowcolor{lightgray!20}
        \multicolumn{1}{|c|}{\textbf{miss}}  & \multirow{-1.65}{*}{\textbf{Name}} & \multirow{-1.65}{*}{\textbf{MPKI}} & \multicolumn{1}{c}{\textbf{inst.}} & \multicolumn{1}{c}{\textbf{inst.}} & \multicolumn{1}{c}{\textbf{(GB)}}          & \multicolumn{1}{c|}{\textbf{hit ratio}}
                                            & \multicolumn{1}{|c|}{\textbf{miss}} & \multirow{-1.65}{*}{\textbf{Name}} & \multirow{-1.65}{*}{\textbf{MPKI}} & \multicolumn{1}{c}{\textbf{inst.}} & \multicolumn{1}{c}{\textbf{inst.}} & \multicolumn{1}{c}{\textbf{(GB)}}          & \multicolumn{1}{c|}{\textbf{hit ratio}} \\ \hline
        \multirow{10}{*}{\textbf{high}}
            & {gcc}      & 3.08 & 28.0\% & 14.0\% & 1.0 & 20.1\%
            & \multirow{10}{*}{\textbf{low}} & {hmmer}    & 1.17 & 12.9\% & 16.3\% & 1.1 & 22.1\% \\ \cline{2-7}\cline{9-14}
            & {gobmk}    & 2.38 & 16.4\% & 13.8\% & 2.0 & 29.9\%
            &            & {leslie3d} & 1.16 & 14.5\% & 18.5\% & 12.7 & 27.4\% \\ \cline{2-7}\cline{9-14}
            & {cactus}   & 2.37 & 23.5\% & 17.8\% & 22.7 & 17.1\%
            &            & {quantum}  & 1.08 & 21.2\% & 19.5\% & 6.8  & 25.7\% \\ \cline{2-7}\cline{9-14}
            & {milc}     & 1.60 & 21.2\% & 19.0\% & 34.9  & 21.6\%
            &            & {AES}      & 1.07 & 8.8\%  & 14.6\% & 6.9  & 21.1\% \\ \cline{2-7}\cline{9-14}
            & {bzip2}    & 1.50 & 21.7\% & 19.7\% & 96.0 & 24.2\%
            &            & {astar}    & 0.88 & 24.3\% & 17.1\% & 0.3  & 25.4\% \\ \cline{2-7}\cline{9-14}
            & {lbm}      & 1.35 & 16.4\% & 38.5\% & 42.2 & 24.3\%
            &            & {SHA512}   & 0.86 & 11.7\% & 4.5\%  & 0.3  & 22.2\% \\ \cline{2-7}\cline{9-14}
            & {sjeng}    & 1.32 & 15.9\% & 38.5\% & 17.8  & 26.3\%
            &            & {calculix} & 0.80 & 26.6\% & 15.2\% & 1.0  & 15.9\% \\ \cline{2-7}\cline{9-14}
            & {namd}     & 1.31 & 22.1\% & 14.8\% & 6.0 & 25.7\%
            &            & {povray}   & 0.55 & 30.4\% & 12.7\% & 0.2  & 4.2\%  \\ \cline{2-7}\cline{9-14}
            & -          & -    & -      & -      & -          & -
            &            & {tonto}    & 0.54 & 16.6\% & 10.9\% & 0.4  & 14.2\% \\ \cline{2-7}\cline{9-14}
            & -          & -    & -      & -      & -          & -
            &            & {bwaves}   & 0.50 & 27.6\% & 9.2\%  & 34.3  & 17.0\% \\ \hline
      \end{tabular}
    }
    \subcaption{Workload characteristics.} \label{tab:workload}
  \end{minipage}
  \vspace{15pt}
  \caption{Evaluation setup.} \label{tab:eval_setup}
  \vspace{5pt}
\end{table*}

\noindent \textbf{Latency and persistence controls.} As shown in Figure \ref{fig:design_extention-annotation}, when memory instructions, such as load (LD) and store (ST), arrive, the CXL RP generates the corresponding hints, which are classified into two categories: \emph{Bufferability} and \emph{Determinism}, detailed below. These features are appended to the reserved area of M2S Req messages, and the endpoint controller forwards them to the underlying SSD controller. By interpreting these annotations, the SSD controller can reschedule internal tasks such as garbage collection or determine whether target data should be cached in its internal DRAM, thereby improving latency and persistence characteristics when integrating CXL with SSDs.

Determinism is defined by two states: deterministic (\textbf{DT}) and non-deterministic (\textbf{ND}). DT signals that the host requires the Type 3 device to handle the tagged request without involving internal tasks, ensuring predictable performance. In contrast, ND allows the corresponding requests to be processed in a fire-and-forget manner, enabling the device to schedule internal tasks either during subsequent ND requests or during idle periods.

We enable the host system to transparently apply annotations regarding determinism to users. SSD tail latency becomes particularly problematic when subsequent instructions depend on specific load instructions. Unlike conventional block read requests for PCIe SSDs, read requests issued to CXL-connected SSDs are memory read requests, inherently limiting the extent of asynchronous handling. In addition, these requests can directly cause CPU pipeline stalls because their latency is significantly longer than typical local memory accesses. Thus, to minimize performance degradation caused by SSD tail latency, it is essential for the CPU to consider the currently executing instructions. However, due to instruction reordering and related techniques, the CPU itself is best suited to evaluate how likely a specific load instruction is to result in pipeline stalls. For this reason, we propose that when the proportion of load instructions in the CPU's instruction queue and reorder buffer exceeds a certain threshold, the CPU should issue CXL.mem requests with the \textbf{DT} annotation at runtime. This method can minimize performance degradation caused by SSD tail latency while simultaneously enhancing user convenience.

Bufferability is similarly composed of two states: bufferable (\textbf{BF}) and non-bufferable (\textbf{NB}). Requests annotated as BF can be cached or buffered in the SSD\textquotesingle{}s internal DRAM, optimizing performance for scenarios where persistence is not immediately required. Conversely, NB requests prioritize persistence, treating it as a first-class requirement. For such cases, the PCIe storage can selectively write requests directly to block media, avoiding the need to globally flush all data residing in its large internal DRAM to the block media at once.

Note that, unlike determinism, which relates exclusively to performance, bufferability depends on application persistence requirements. Therefore, annotations reflecting user intent significantly aid the decision-making of the underlying storage. Since applies only to critical persistence store instructions, we propose enabling CPUs to provide specifically annotated store instructions. Unlike regular store instructions, our store instruction explicitly sends a CXL.mem packet annotated with \textbf{NB} to the CXL-SSD. The SSD then identifies that it should written directly to backend media without buffering. Consequently, when the memory request completes, the host can be certain that the corresponding store has persisted. In database applications, users can apply this store instruction exclusively to memory requests related to transaction journaling, thereby ensuring persistence with minimal overhead.

\noindent \textbf{Details of annotations and examples.} The annotations--determinism, bufferability, and GPF--can be used individually or in combinations (e.g., BF+DT, BF+ND, NB+DT, NB+ND) to address diverse user scenarios. For instance, databases and transactional memory libraries (e.g., \texttt{libpmem} and \texttt{libpmemobj}) often log data when a transaction begins. Since this log data does not need to be persistent until the transaction is committed, the host can annotate these operations with BF+ND or NB+ND. This enables the storage device to buffer incoming writes and allocate time for internal tasks. When the transaction commits, the system can use GPF to flush all buffered data and annotate the commit operation (if necessary) with NB+DT, ensuring persistence within strict latency bounds.

Consider the transaction management of RocksDB as an example. When initiating a transaction, operations can be annotated with ND, as the start of a transaction is not latency-sensitive and does not require immediate persistence. Similarly, query loggings such as Put and Get operations can also benefit from ND for the same reasons. However, when a transaction commit occurs, the associated commit log must be stored persistently within a minimal latency window. By annotating commit commands with NB+DT, the system ensures that the commit log is written to the SSD\textquotesingle{}s persistent storage promptly. With the DT annotation, the SSD halts ongoing internal tasks such as garbage collection, prioritizing the commit log to meet both CXL memory interface characteristics and SSD block storage requirements.

Since most instructions rely on the arrival of their operands, load operations typically benefit from DT. However, if there is no subsequent instruction requiring the result of a previously issued operation (i.e., no read-after-write dependency), synchronization of the current loads is unnecessary. In such cases, annotations like BF+DT or BF+ND can be used for load operations, enabling the storage to prefetch data into its internal DRAM. For instance, in loop-based code segments (e.g., matrix computations), where data exhibits spatial and temporal locality, these annotations inform the storage that the data is likely to be accessed repeatedly, ensuring it is readily available in internal DRAM for future operations.

Another scenario where these annotations are advantageous is lock and synchronization management. Mechanisms such as spinlocks, fences, and barriers typically do not require persistence but are highly sensitive to latency. For example, a spinlock often uses atomic instructions like compare-and-swap or compare-and-exchange, which involve both a memory read and write. Spinlocks generally avoid placing their parameters in the CPU cache, causing the atomic instruction to repeatedly access the same memory address. In this case, annotating both load and store operations with BF+DT ensures low-latency access to Type 3 devices. Similarly, memory fences and barriers benefit from BF+DT annotations as their scope is typically confined to the running process rather than requiring long-term persistence of data. These mechanisms prioritize immediate responsiveness, making bufferability and determinism well-suited for such use cases.

%% file: evaluation.tex
\begin{figure*}
  \includegraphics[width=1\linewidth]{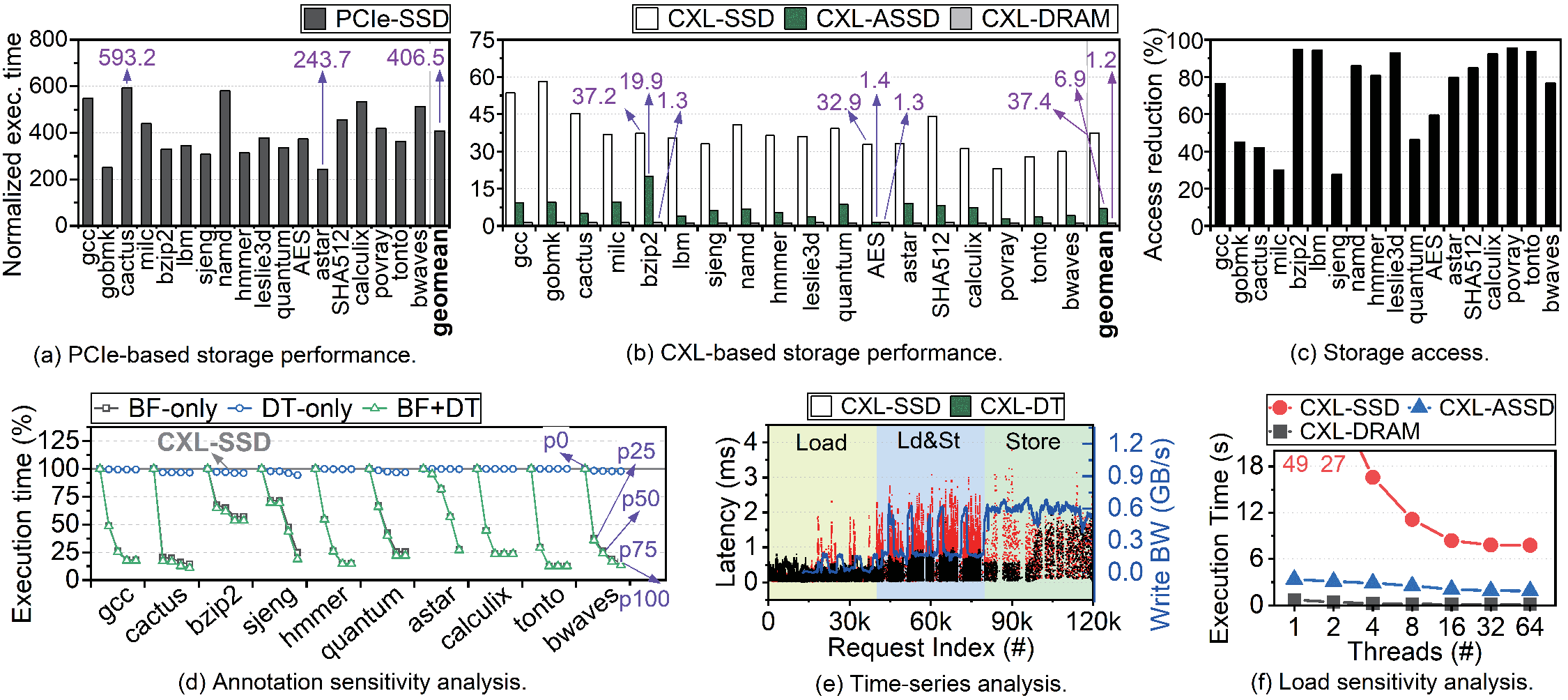}
  \vspace{0pt}
  \caption{Performance analysis.}
  \label{fig:evaluation}
  \vspace{5pt}
\end{figure*}

\subsection{Evaluation Setup}
\noindent \textbf{Methodology.}
To explore a complete design space, we use a full system simulation model that covers from the CPU to the block storage.
Specifically, the system combines gem5 \cite{binkert2011gem5} and SimpleSSD \cite{gouk2018amber} for CXL RP and CXL-enabled SSD, respectively, by modifying them with the actual cycles we observed in the Preliminary Performance Model of CXL-SSD section.
The main configuration parameters that we used for the evaluations are explained Table \ref{tab:setup}.

\noindent \textbf{Workloads.} We evaluate eighteen workloads sourced from SPEC CPU \cite{henning2006spec} and RV8 Bench \cite{rv8bench}, representing a diverse mix of computation-intensive applications (e.g., encryption algorithms and HPC tasks) and memory-intensive behaviors (e.g., in-memory databases).
To characterize these workloads, we analyze their last-level cache (LLC) hit ratios, LLC misses per thousand instructions (MPKI), and load/store ratios, as detailed in Table \ref{tab:workload}.
The workloads are sorted in descending order of LLC MPKI; workloads with an MPKI greater than 1.2 are identified as exhibiting high cache miss rates.

\noindent \textbf{Configurations.}
We evaluate three memory expansion systems (\texttt{PCIe-SSD}, \texttt{CXL-SSD}, and \texttt{CXL-ASSD}) and compare them against two ideal baseline systems (\texttt{DRAM} and \texttt{CXL-DRAM}).
All three expansion systems (\texttt{PCIe-SSD}, \texttt{CXL-SSD}, and \texttt{CXL-ASSD}) utilize the same underlying media for memory expansion. However, their interfaces differ: \texttt{PCIe-SSD} exposes internal memory solely through PCIe BAR, making it non-cacheable; \texttt{CXL-SSD} employs CXL as the media interface, mapping the memory to the host\textquotesingle{}s cacheable system memory; and \texttt{CXL-ASSD} integrates internal DRAM management and task scheduling, leveraging bufferability and determinism.

For baseline comparisons, \texttt{DRAM} represents DRAM directly attached to the CPU via a conventional DDR interface, while \texttt{CXL-DRAM} connects via CXL instead of DDR but exclusively uses DRAM as its backing media.

\subsection{Performance Analysis}
Figures \ref{fig:evaluation}\textcolor{blue}{a} and \ref{fig:evaluation}\textcolor{blue}{b} present an analysis of the execution time behaviors of PCIe-based storage and three CXL-based storage configurations.
To facilitate comparison, all results are normalized to the execution time of \texttt{DRAM}.

\noindent \textbf{PCIe-based storage.}
\texttt{PCIe-SSD} demonstrates a 406.5$\times$ longer execution time compared to \texttt{DRAM}, on average. This performance gap arises primarily from two factors. First, the SSD controller and flash memory in PCIe-SSDs are optimized for block-based interfaces, which operate at coarse granularity and significantly degrade performance in memory-intensive operations. Second, every load/store memory request from the host is routed directly to the PCIe Base Address Register (BAR), rendering the benefits of on-chip caching ineffective. This is evident in the fact that \texttt{PCIe-SSD} performance remains unaffected by MPKI variations.

The workload-dependent performance variations are mainly influenced by the extent to which the internal DRAM cache can be utilized. For instance, load-intensive workloads (e.g., cactus) experience the worst performance because the CPU is stalled until the requested data is fully loaded into its registers. In contrast, store-intensive workloads (e.g., astar) suffer less performance degradation since their data can be temporarily buffered in the internal DRAM, allowing the CPU to continue operations without significant interruptions. Nevertheless, \texttt{PCIe-SSD} still struggles with accesses to non-cacheable regions, which further impacts its overall performance.

\noindent \textbf{CXL-based storage.}
Figure \ref{fig:evaluation}\textcolor{blue}{b} compares the performance of \texttt{CXL-SSD}, \texttt{CXL-ASSD}, and \texttt{CXL-DRAM} against \texttt{DRAM}.
As anticipated, \texttt{CXL-SSD} achieves an average performance improvement of 10.9$\times$ over \texttt{PCIe-SSD}.
Although the underlying storage media and characteristics remain identical, \texttt{CXL-SSD} benefits from its placement in cacheable memory space via CXL.
This placement significantly reduces storage access frequency by 72.1\% (see Figure \ref{fig:evaluation}\textcolor{blue}{c}), resulting in notable performance gains.

The presence of workload locality, despite variations, further enhances the performance of \texttt{CXL-SSD}.
Specifically, workloads with low cache miss rates (e.g., bwaves, tonto, povray) reduce storage accesses by 80.2\%, yielding an average performance improvement of 12.2$\times$.
In contrast, workloads with high cache miss rates reduce storage accesses by 62.1\%, leading to an average improvement of 9.6$\times$.
These findings underscore the importance of leveraging CXL\textquotesingle{}s cacheable memory architecture to mitigate storage access overhead and improve overall execution efficiency.

\texttt{CXL-ASSD} achieves an average performance improvement of 5.4$\times$ compared to \texttt{CXL-SSD}.
While the total number of storage accesses remains unchanged between \texttt{CXL-ASSD} and \texttt{CXL-SSD}, the BF/NB and DT/ND annotations allow the underlying CXL and SSD controllers to better utilize internal DRAM.
This optimization effectively hides the long latency associated with flash media.

Note that \texttt{CXL-DRAM} demonstrates performance approaching that of \texttt{DRAM}.
Although \texttt{CXL-DRAM} relies on a slightly slower PCIe interface compared to high-speed DDR, the inherent locality of most applications and the benefits of on-chip caching help mitigate this disadvantage.
While \texttt{CXL-ASSD} may exhibit less-comparable performance for few workloads such as bzip2, \texttt{CXL-ASSD} has the potential to close the perfomance gap. We will also explore this with details shortly.

\noindent \textbf{Sensitivity Tests.}
{
    While previously discussed \texttt{CXL-ASSD} applies annotations for all the commonly used functions, depending on the environment, one might not be feasible to annotate as such.
    Fortunately, we found that even with less amount of annotations, \texttt{CXL-ASSD} still exhibit comparable performance.
    Specifically, we evaluated \texttt{CXL-ASSD} with varying amount of annotations applied.
    In addition, we also evaluated how different kinds of annotation (bufferability and determinism) affects the overall performance.
    For DT, we adjusted the threshold of load instructions in the instruction queue/reorder buffer at which the CPU begins applying deterministic (\textbf{DT}) annotations, thereby annotating a specific percentage of total instructions. For BF, we varied the proportion of functions annotated as bufferable (\textbf{BF}).

    The results are shown in Figure \ref{fig:evaluation}\textcolor{blue}{d}.
    The y-axis shows the execution time, normalized to that of \texttt{CXL-SSD}.
    The connected five dots for a workload describes the execution time when different amount of annotation are applied (0\%, 25\%, 50\%, 75\%, 100\%, respectively).
    BF-only and DT-only, as the name suggest, only applies certain kinds of annotations while BF+DT applies both annotations.
    The \texttt{pN} labels (e.g., p25, p50, etc.) represent the performance achieved when $N$\% of the total instructions (for DT) or functions (for BF) are annotated.

    As shown, annotating only 25\% of commonly used functions reduces average execution time by 50.1\%. This occurs because few functions dominate storage accesses; for instance, just eight functions (e.g., \texttt{\_raw\_spin\_lock}) account for 50.5\% of total storage accesses.  Though workloads like \textit{bzip2}  initially show limited performance gains, additional per-application analyses can effectively address these cases. Specifically annotating \texttt{clear\_page\_erms}, the primary storage access source for \textit{bzip2}, reduces its execution time by 77.4\%.

\begin{figure*}
    \begin{subfigure}{\linewidth}
        \vspace{10pt}
        \centering
        \includegraphics[width=1\linewidth]{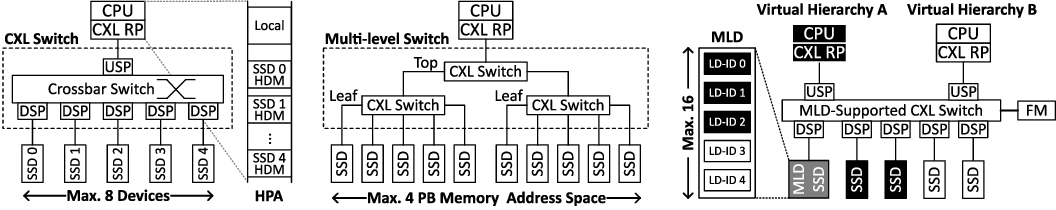}
        \begin{tabularx}{\textwidth}{
            p{\dimexpr.28\linewidth-2\tabcolsep-1.3333\arrayrulewidth}
            p{\dimexpr.36\linewidth-2\tabcolsep-1.3333\arrayrulewidth}
            p{\dimexpr.38\linewidth-2\tabcolsep-1.3333\arrayrulewidth}
            }
              \vspace{-5pt} \caption{Switch.} \label{fig:disaggregation-switch}
            & \vspace{-5pt} \caption{Multi-level Switch.} \label{fig:disaggregation-multi-switch}
            & \vspace{-5pt} \caption{Virtual hierarchy.} \label{fig:disaggregation-mld}            
        \end{tabularx}
    \end{subfigure} 
    \vspace{0pt}
    \caption{Storage disaggregation configurations.} \label{fig:disaggregation}
    \vspace{5pt}
\end{figure*}

    The results further indicate BF annotations contribute most significantly enhance \texttt{CXL-ASSD} performance. This is because BF-only configurations consistently leverage the internal DRAM of SSDs to reduce latency, while DT-only configurations mitigate long-tail latency issue. Nevertheless, DT annotations remain essential for maintaining stability, as tail latency occasionally reaches several milliseconds. Combining BF and DT annotations provides a balanced approach, enhancing overall efficiency and reliability.

    Figure \ref{fig:evaluation}\textcolor{blue}{e} shows how DT annotations migrate SSD tail latency. To evaluate this, we compared a load-intensive workload, \texttt{cactus}, under two conditions: without DT annotations (\texttt{CXL-SSD}) and with DT annotations (\texttt{CXL-DT}). The workload \texttt{cactus} initially exhibits a load-intensive behavior, gradually becoming store-intensive pattern over time. Memory write bandwidth over time is indicated by the right y-axis (blue line). For this evaluation, the CPU configuration matched that of the \texttt{p75} scenario in Figure \ref{fig:evaluation}\textcolor{blue}{d}, meaning the threshold was set to annotate approximately 75\% of all instructions with \textbf{DT}. The figure shows memory latency experienced by each load/store instruction, presented in execution order.

    The evaluation shows \texttt{CXL-SSD} experiences prolonged tail latencies from random SSD-internal tasks, even with a high proportion of load instructions. These millisecond-scale memory delays cause previously unseen, prolonged CPU pipeline stalls, significantly degrading performance. In contrast, \texttt{CXL-DT} avoids such issues by enabling the CPU to monitor the proportion of load instructions, providing hints to suppress internal SSD activities. As a result, \texttt{CXL-DT} successfully avoids tail latency throughout the entire load-intensive period (until 90k-th instruction). However, DT annotations cannot fundamentally eliminate SSD tasks. As the application transitions to a store-intensive pattern around the 90k-th instruction, the CPU stops sending DT annotations, allowing internal SSD tasks to resume, causing long tail latencies to reappear. Thus, although \texttt{CXL-DT} still encounters long tail latencies, these occur only when the application is store-intensive. Under these conditions, the CPU effectively avoid performance degradation due to pipeline stalls, preserving the efficiency of the fire-and-forget policy.

    On the other hand, Figure \ref{fig:evaluation}\textcolor{blue}{f} illustrates the latency and bandwidth characteristics of CXL memory expansion according to workload type. For evaluation, we utilized the STREAM microbenchmark suite \cite{McCalpin1995}, with vector kernels (copy, scale, add, triad), and varied  host threads across three configurations: \texttt{CXL-SSD}, \texttt{CXL-ASSD}, and \texttt{CXL-DRAM}. For the \texttt{CXL-ASSD}, we adopted the same \texttt{p75} settings used in Figure \ref{fig:evaluation}\textcolor{blue}{d}, applying both BF and DT annotations. The results show that as the number of workload threads decreases, the performance gap between \texttt{CXL-SSD} and \texttt{CXL-ASSD} widens. This occurs because fewer threads become more sensitive to latency rather than bandwidth provided by CXL memory. Consequently, by leveraging the SSD's internal DRAM, \texttt{CXL-ASSD} achieves up to 14.6$\times$ higher performance compared to \texttt{CXL-SSD}.

    In contrast, as the thread count increases, application performance relies more on CXL memory bandwidth, reducing performance gap between \texttt{CXL-SSD} and \texttt{CXL-ASSD}. Nevertheless, even with 64 threads, \texttt{CXL-ASSD} still achieves a 4.2$\times$ performance improvement over \texttt{CXL-SSD}. This is because the internal DRAM within the SSD provides significantly higher bandwidth than Z-NAND. Thus, despite being Z-NAND-based memory expansion, \texttt{CXL-ASSD} performance is 4.7$\times$ lower than \texttt{CXL-DRAM}.
}

%% file: design_disaggregation.tex
This section discusses how a system can disaggregate CXL controllers and storage devices from its computing resources while keeping their byte-addressability.

\noindent\textbf{Pooling storage over the byte interface.} To make the interconnect network scalable, CXL 3.1 allows FlexBus to employ one or more CXL switches, each being able to have multiple upstream ports (USPs) and downstream ports (DSPs). Even though CXL yet leaves a question undecided on how to implement a switch and its internal components, USPs and DSPs can be simply interconnected by a reconfigurable crossbar switch. Specifically, a USP can be connected to a CXL RP or another switch\textquotesingle{}s DSP over FlexBus, and it internally returns the incoming message to one or more underlying DSPs as soon as possible. In contrast, a DSP links a lower-level hardware module such as a storage device\textquotesingle{}s CXL endpoint or a different switch\textquotesingle{}s USP. Using a switch buffer, it can control multiple CXL messages going through the DSP(s).

Figure \ref{fig:disaggregation-switch} shows how a host can expand its local mem-
ory by having multiple storage devices. Specifically, each DSP connects to a different storage device, whereas a USP is linked to all the DSPs and exposes them to the host\textquotesingle{}s CXL RP. For this type of storage-integrated memory expansion, the host should map each HDM to different places of its physical memory and be aware of the mapping information when the system enumerates CXL devices to configure CXL capability/configuration (BAR/HDM). While this network topology is simple enough to connect multiple PCIe storage devices, the number of lanes that a CXL switch can support is limited. Typically, a switch supports 64$\sim$128 lanes, and thus, only 4$\sim$8 ports are available for high-performance storage devices (using 16 lanes). In this case, as shown in Figure \ref{fig:disaggregation-multi-switch}, it can expand the host memory by adding one or more switches to the CXL network. The top switch is used to bridge the host\textquotesingle{}s CXL RP and all other lower-level switches, while the leaf switches are employed to manage many PCIe storage devices. Note that the number of storage devices that a network can handle varies based on the size of the devices and the memory capacity that CXL deals with (currently, it is 4PB).

\noindent \textbf{Multi-host connection management.} To better utilize the
storage resources, we can also connect arbitrary numbers of host CPUs to the CXL network. Since the switch\textquotesingle{}s crossbar (called fabric manager) remembers each connection between USPs and DSPs, we can fabricate a unique routing path beginning from a host to one or more storage devices, called virtual hierarchy (VH). Each VH guarantees that a storage device can be mapped to a host, which is attached anywhere in the CXL network. Thus, VHs allow the system to completely disaggregate many PCIe storage devices from its multi-host computing resources for the memory expansion. While these reconfigurable VHs can realize a fully scale-out architecture, memory resources expanded by the storage devices are unfortunately tricky to control finely. Since the storage device should only be associated with a host, it can be underutilized and/or unbalanced across different CPUs.

\noindent\textbf{Storage device virtualization.} To address this issue, we can virtualize each storage device to be shared by different hosts. Specifically, as shown in Figure \ref{fig:disaggregation-mld}, CXL allows a system to logically splits each endpoint into multiple Type 3 devices (up to 16), called multiple logical device (MLD). Thus, we can make each MLD define its own HDM, which can be mapped to a different place of any host of system memory, similar to a physical storage device. As each MLD associated with the same storage device can be a part of different VHs, it is expected to utilize the underlying storage resources better by allocating the memory expanders in a fine granular manner.

A disadvantage of multi-host VHs can be bandwidth sharing and/or traffic congestion. To support MLDs, PCIe storage may require partitioning the underlying backend and internal DRAMs, lowering the level of parallelism, and this can unfortunately reduce the bandwidth of each MLD. In addition, as a single storage device (and a CXL switch) can be shared by multiple hosts, the endpoint\textquotesingle{}s fabric can be congested more than before. Since the performance of this multi-host memory expansion varies based on diverse perspectives and hardware configurations of CXL, it does need careful network and storage designs.

%% file: conclusion.tex
This paper explores how CXL can transform PCIe-based block storage into scalable, byte-addressable working memory. By leveraging CXL\textquotesingle{}s cacheability and adopting Type 3 endpoint devices (CXL-SSDs), we bridge block-based PCIe semantics with memory-compatible operations. To narrow the performance gap with DRAM, we propose annotation mechanisms, Determinism and Bufferability, to improve performance while maintaining data persistence. Our FPGA-based prototype and simulation demonstrate that CXL-SSDs significantly outperform PCIe-based memory expanders and approach DRAM-like performance in workloads with high locality, demonstrating the feasibility of integrating block storage into CXL\textquotesingle{}s memory ecosystem.

%% file: acknowledgements.tex
This work is protected by one or more patents.
The authors would like to thank the anonymous
reviewers for their comments, and Myoungsoo Jung is the
corresponding author (mj@panmnesia.com).